\documentclass[12pt]{article}

\usepackage{epsfig}
\usepackage{amssymb}
\usepackage{subfigure}
\usepackage{graphicx}
\usepackage{color}
\usepackage{jheppub} 

\begin{document}

\baselineskip 6mm
\renewcommand{\thefootnote}{\fnsymbol{footnote}}


\newcommand{\rnc}{\renewcommand}



\newcommand{\tcb}{\textcolor{blue}}
\newcommand{\tcr}{\textcolor{red}}
\newcommand{\tcg}{\textcolor{green}}


\def\be{\begin{eqnarray}}
\def\ee{\end{eqnarray}}
\def\nn{\nonumber\\}


\def\ct{\cite}
\def\la{\label}
\def\eq#1{\eqref{#1}}


\def\a{\alpha}
\def\b{\beta}
\def\g{\gamma}
\def\G{\Gamma}
\def\d{\delta}
\def\D{\Delta}
\def\e{\epsilon}
\def\et{\eta}
\def\ph{\phi}
\def\Ph{\Phi}
\def\ps{\psi}
\def\Ps{\Psi}
\def\k{\kappa}
\def\l{\lambda}
\def\L{\Lambda}
\def\m{\mu}
\def\n{\nu}
\def\th{\theta}
\def\Th{\Theta}
\def\r{\rho}
\def\s{\sigma}
\def\S{\Sigma}
\def\ta{\tau}
\def\o{\omega}
\def\O{\Omega}
\def\pr{\prime}


\def\half{\frac{1}{2}}
\def\goto{\rightarrow}

\def\na{\nabla}
\def\grad{\nabla}
\def\curl{\nabla\times}
\def\div{\nabla\cdot}
\def\pa{\partial}
\def\fr{\frac}

\def\bra{\left\langle}
\def\ket{\right\rangle}
\def\lb{\left[}
\def\lc{\left\{}
\def\ls{\left(}
\def\lp{\left.}
\def\rp{\right.}
\def\rb{\right]}
\def\rc{\right\}}
\def\rs{\right)}

\def\vac#1{\mid #1 \rangle}


\def\td#1{\tilde{#1}}
\def\check{ \maltese {\bf Check!}}


\def\Tr{{\rm Tr}\,}
\def\det{{\rm det}}
\def\text#1{{\rm #1}}


\def\bc#1{\nnindent {\bf $\bullet$ #1} \\ }
\def\ch {$<Check!>$ }
\def\ss {\vspace{1.5cm}}
\def\inf{\infty}

\begin{titlepage}

\hfill\parbox{2cm} { }

\hspace{13cm} \today
 
\vspace{2cm}

\begin{center}
{\Large \bf Holographic RG flow \\
triggered by a classically marginal operator}

\vskip 1. cm
   {Chanyong Park$^{a}$\footnote{e-mail : cyong21@gist.ac.kr}}

\vskip 0.5cm

{\it $^a$ Department of Physics and Photon Science, Gwangju Institute of Science and Technology,  Gwangju  61005, Korea}

\end{center}

\thispagestyle{empty}

\vskip3cm


\centerline{\bf ABSTRACT} 


\vspace{1cm}

We study the holographic renormalization group (RG) flow triggered by a classically marginal operator. When a marginal operator deforms a conformal field theory, it does not yield a nontrivial renormalization group flow at the classical level. At the quantum level, however, quantum corrections modify a marginal operator into one of the truly marginal, marginally relevant, and marginally irrelevant operators and can generate a nontrivial RG flow. We investigate the holographic description of a RG flow triggered by a marginal operator with quantum corrections. We look into how the physical quantities of a deformed theory, a coupling constant and the vacuum expectation value, rely on the RG scale. We further discuss the holographic description of the trace anomaly caused by the gluon condensation.

\vspace{2cm}

\end{titlepage}

\renewcommand{\thefootnote}{\arabic{footnote}}
\setcounter{footnote}{0}

\tableofcontents


\section{Introduction}

For the last decade, much attention has been paid to understanding strongly interacting systems of Quantum Chromodynamics (QCD) and condensed matter theory by using the anti de-Sitter (AdS)/conformal field theory (CFT) correspondence \cite{Maldacena:1997re,Gubser:1998bc,Witten:1998qj,Witten:1998zw}. The AdS/CFT correspondence has passed many nontrivial checks. When a relevant operator deforms a CFT, the original CFT is modified  relying on the energy scale observing the system and eventually flows a new IR theory \cite{Wilson:1973jj}. At a low energy scale, systems of nuclear and condensed matter physics usually have a strong coupling constant. Understanding nonperturbative physics governed by strong interactions is one of the important issues in recent theoretical and experimental researches. To account for such nonperturbative phenomena, we need to figure out an exact and nonperturbative renormalization group (RG) flow. Although the perturbative RG flow was well established in the quantum field theory (QFT), it remains an important issue to understand the nonperturbative RG flow. The AdS/CFT correspondence or holography can shed light on this issue due to the  relation between the nonperturbative RG flow of the QFT and the equations of motion of the dual gravity \cite{Henningson1998,Henningson2000,Freedman1999,Gubser:1999vj,deBoer:1999tgo,Verlinde:1999xm,Boer2001,Skenderis:1999mm,Polchinski:2011im,Papadimitriou:2010as,Papadimitriou:2004ap,Skenderis2002,Heemskerk:2010hk,Lee:2017nma}. In this work, we investigate the holographic RG flow of a dual QFT deformed by a marginal operator and compare the holographic result with the gluon condensate appearing in the lattice QCD \cite{DiGiacomo:1981lcx,Trinchero:2013joa}.

After the AdS/CFT conjecture, there were many attempts to understand the dual QFT of an asymptotic AdS geometry. In the holographic setup, an IR energy scale of the gravity theory maps to a UV energy scale of the dual field theory and generally leads to various UV divergences \cite{Gubser:1999vj,deBoer:1999tgo,Verlinde:1999xm,Skenderis:1999mm}. In order to get rid of such UV divergences, several distinct methods have been invented. One of them is the `background subtraction' method, which gets rid of the effect of the background reference spacetime. This subtraction method is not applicable to certain cases where an appropriate reference solution is ambiguous or unknown, e.g., topological black holes \cite{Mann:1996gj,Brill:1997mf,Vanzo:1997gw,Emparan:1998he,Birmingham:1998nr}, Taub-NUT-AdS, and Taub-bolt-AdS \cite{Chamblin:1998pz,Emparan:1999pm,Hawking:1998ct}. Another way to remove the UV divergence is called the holographic renormalization in which the UV divergences are removed by adding appropriate counterterms. This method is similar to the renormalization process of a QFT. The holographic renormalization method is useful in that it does not require a background reference spacetime \cite{deBoer:1999tgo}. On the dual field theory side, the holographic renormalization corresponds to the momentum space RG flow. There is another description called the RG flow of the entanglement entropy where the RG scale is characterized by the subsystem size rather than the boundary position. It has been shown that the thermodynamics-like law of the entanglement entropy can reproduce the linearized Einstein equation of the dual AdS geometry. From the field theory point of view, the RG flow of the entanglement entropy corresponds to the real space RG flow \cite{Ryu:2006bv,Ryu:2006ef,Nishioka:2009un,Takayanagi:2012kg,Narayanan:2018ilr,Park:2018ebm,Park:2019pzo}. Intriguingly, it was shown by applying the entanglement entropy RG flow that a quantum entanglement entropy in the UV region can flow to a thermal entropy in the IR region \cite{Kim:2016jwu}. From now on, we concentrate on the momentum space RG flow to find the connection between the gravity equations and the RG equations.

When one applies the holographic renormalization technique to the gravity with some matter fields, it was shown that there can exist an ambiguity in choosing the counterterms \cite{Park:2013dqa,Park:2014raa}. In the holographic Lifshitz theory, for example, there are many different counterterms giving the same UV divergence.  Although the dual field theory is uniquely determined at a UV fixed point, choosing different counterterm leads to a different RG flow. This implies that, although we start with the same UV theory, the low energy physics along the RG flow can show different physics relying on the counterterm we took. Therefore, it is important to clarify the counterterm for describing the IR physics correctly in terms of the RG flow. In this work, we investigate how we can get information about the RG flow after the holographic renormalization with specifying the counterterms. To do so, we exploit the Hamilton-Jacobi formalism of the gravity theory \cite{deBoer:1999tgo,Verlinde:1999xm,Boer2001}. After finding the relation between the radial coordinate and the RG scale, we show how the Hamilton-Jacobi formalism reproduces to the RG equation of the dual QFT.

When a CFT deforms by a marginal operator, the deformed theory remains marginal at the classical or action level. However, this not true at the quantum level because quantum corrections may change the property of the deformation operator. The AdS/CFT correspondence claimed that a classical gravity theory is dual to a full quantum field theory defined at the boundary. Therefore, the holographic description allows us to figure out the nonperturbative RG flow containing the quantum effect. One of the important quantities affected by such quantum corrections is the $\b$-function. The $\b$-function describes how the coupling constant depends on the energy scale. In general, physical phenomena crucially depends on the strength of a coupling constant. At the classical level, the $\b$-function can be classified by the conformal dimension of a deformation operator. When the coupling constant $\l$ couples to an operator with the conformal dimension $\D$ in a $d$-dimensional CFT, the classical $\b$-function is given by \cite{Hollowood:2009eh}
\be
\b_{cl} = - (d -\D) \l .
\ee
If we further take into account quantum corrections, the $\b$-function generalizes to
\be
\b = \b_{cl} + \b_{q}  ,
\ee
where $\b_{q}$ indicates quantum corrections. If we concentrate on a classically marginal operator with $\D=d$, the classical $\b$-function automatically vanishes. However, the quantum effect leads to a nonvanishing $\b$-function and generates a nontrivial RG flow. As explained here, the quantum correction plays a crucial role in determining the RG flow triggered by a marginal operator. In this work, we investigate the holographic description of such quantum correction and discuss how such a marginal deformation modifies a UV CFT.

Related to the marginal deformation in QCD, one of the important quantities is the gluon condensation, $\bra G \ket $. Its one-loop correction in the UV region leads to the following trace anomaly \cite{DiGiacomo:1981lcx,Trinchero:2013joa}
\be
\bra {T^{\m}}_{\m} \ket = - \fr{N_c}{8 \pi} \fr{\b_\l}{\l^2} \bra  G \ket ,
\ee
where $\l$ and $\b_\l$ indicate the t' Hooft coupling and its $\b$-function, respectively. This trace anomaly is crucially associated with the RG flow caused by the gluon condensation. In the present work, we investigate the trace anomaly and  its RG scale dependence by applying the holographic RG flow technique. 

The rest of this paper is organized as follows. In Sec. 2, we discuss the $\b$-function of a marginal operator and the RG equation appearing in a QFT. In Sec. 3, we study a gravity theory involving a scalar field which is dual of a marginal operator. In addition, we investigate how to construct the holographic RG flow caused by such a marginal operator with discussing the correct counterterms required to renormalize the UV divergences. By applying this holographic RG flow technique, we investigate the trace anomaly caused by the gluon condensation in Sec. 4. We finish this work with some concluding remarks in Sec. 5.


\section{RG flow in QFT}

At a low energy scale, understanding new physics laws of macroscopic systems is one of the important issues to be resolved. To understand macroscopic theories, we need to know how a microscopic UV theory flows to a new IR theory along the RG flow. A relevant or marginally relevant operator causes a nontrivial RG flow and its effect becomes signigicant in the IR region. In this RG flow process, nonperturbative quantum corrections play a crucial role. Unfortunately, we do not have an analytic method accounting for the nonperturbative RG flow in the traditional QFT. In this situation, the AdS/CFT correspondence may provide a new chance to investigate such a nonperturbative RG flow.

Before studying the holographic RG flow, we first discuss general aspects of the RG flow in a $d$-dimensional QFT for the comparison with the holographic result. Assuming that a CFT deforms by an operator of a conformal dimension $\D$, the deformed theory is described by
\be
S_{QFT} = S_{CFT}+ \int d^d x \sqrt{\g} \, \m^{d-\D} \l \, \bar{O},
\ee
where $\g_{\m\n}$ corresponds to the background metric  and $\bar{O}$ is a dimensionful composite operator of the fundamental field. Here, $\m$ and  $\l$ denotes an RG scale and a dimensionless coupling constant, respectively. Ignoring quantum effects, the deformation can be characterized by the conformal dimension $\D$. At the classical level,  for example, the dimensionless coupling and the dimensionless operator, $O = \m^{-\D} \bar{O}$,  under the RG or scale transformation scale as \cite{Wilson:1973jj,Hollowood:2009eh}
\be
\l \to \m^{- (d - \D)} \l  \quad {\rm and} \quad  O \to \m^{ - \D} O .   \la{Result:scbehavior}
\ee

Under the RG transformation, the deformation operator is classified into three different categories: relevant for $\D<d$, marginal for $\D=d$, and irrelevant for $\D>d$. The deformation effect of a relevant operator is negligible in the UV regime, while it becomes important in the IR regime and changes the UV theory into another IR theory. A marginal operator at the classical level does not break the conformal symmetry, so that the deformed theory remains conformal in the entire energy range. The irrelevant operator gives rise to a serious effect on the UV theory, so that the resulting theory even in the UV limit differs from the original UV CFT. This means that the deformed theory is UV incomplete. These features become manifest when we take into account a classical $\b$-function \cite{Hollowood:2009eh}
\be
\b_{cl} \equiv \fr{\pa \l}{\pa \log \m} = - (d-\D) \l  .
\ee
The coupling constant determined by the classical $\b$-function automatically satisfies the scaling behavior in \eq{Result:scbehavior}. For a marginal operator with $\D=d$, the classical $\b$-function automatically vanishes and the coupling constant does not change along the RG flow. The coupling constant of an irrelevant operator has a positive $\b$-function, so that the coupling constant becomes large as $\m$ increases. This implies that the effect of the irrelevant deformation becomes more important in the UV region. As a consequence, the deformed theory is not the same as the undeformed one even in the UV limit. On the other hand, a relevant operator has a negative $\b$-function and its coupling constant decreases as $\m$ increases. Therefore, the effect of a relevant operator is negligible in the UV region. Note that this is the story at the classical level. If we further take into account quantum effects, they can modify behaviors of the classical $\b$-function. From now on, we consider a QFT defined at a flat spacetime to ignore the effect of a conformal anomaly.

Quantum effects in the renormalization procedure usually cause UV divergences. Since physical quantities must be finite, we need to get rid of such UV divergences by adding appropriate counterterms. After an appropriate renormalization, the renormalized partition function reduces to
\be
{\cal Z} = \int {\cal D} \ph \, e^{- ( S_{QFT} + S_{ct} ) } = e^{-\G [\g_{\m\n}(\m) ,\l (\m) ;\m]} ,
\ee
where $\G$ is called the generating functional. The background metric $\g_{\m\n}$ in a QFT  usually does not vary along the RG flow. In this work, however, we take into account the background metric as another coupling constant depending the RG scale as done in Ref. \cite{Friedan:1980aa,FRIEDAN1985318}. Since the renormalized partition function is independent of the RG scale, the variation of the partition function gives rise to the following RG equation
\be				\la{Result:gRGeq}
0=  \fr{\m}{\sqrt{\g}} \fr{\pa \G}{\pa \m}  + \g^{\m\n} \bra T_{\m\n} \ket   +   \b_\l  \bra O \ket  ,
\ee
where
\be
\b_\l &=& \fr{d \l}{d \log \m} , \\
\bra T_{\m\n} \ket  &=& -  \fr{2  }{\sqrt{\g}} \fr{\pa \G }{\pa \g^{\m\n}} , \\
\bra O  \ket &=&   \fr{1}{\sqrt{\g}} \fr{\pa \G }{\pa \l}   \la{Relation:vevfromG}.
\ee
When the background metric is independent of the RG scale, the above RG equation reduces to 
\be
0 =  \m \fr{\pa \G}{\pa \m}  +  \b_\l \bra O  \ket    .
\ee
This is the typical RG equation of a QFT. However, assuming that the metric scales under the RG transformation, the generalized RG equation \eq{Result:gRGeq} gives us more information about the stress tensor of a system which plays an important role in understanding the connection to the dual gravity, as will be seen later.  

For the undeformed CFT with $\b_\l=0$, the RG equation reduces to
\be
\bra {T^\m}_{\m} \ket = - \fr{\m}{\sqrt{\g}} \fr{\pa \G}{\pa  \m} . 
\ee
The stress tensor of a CFT must be traceless up to a conformal anomaly \cite{Balasubramanian:1999re}. We may associate the right hand side with a conformal anomaly depending on the topology of the background spacetime. Hereafter, we concentrate on a flat spacetime which has no conformal anomaly. When a CFT deforms by an operator, a nontrivial trace of the stress tensor newly appears and generally breaks the conformal symmetry. Ignoring the RG flow of the generating functional with fixed coupling constants ($\pa \G /\pa \m =0$), the trace of the stress tensor is determined by the deformation
\be			
 \bra {T^\m}_{\m}  \ket   \sim -  \b_\l  \bra O \ket  .
\ee
To avoid the confusion with the conformal anomaly, from now on, we call a nonvanishing  $\bra {T^\m}_{\m}  \ket$ caused by a deformation a trace anomaly.

The quantum effect involved in the renormalization procedure modifies the classical $\b$-function, as mentioned before. For a marginal and a relevant deformation, the $\b$-function in the UV region can be written as  
\be
\b_\l = \b_{cl} + \b_q = - (d-\D) \l  + \b_q  .
\ee
where $\b_q$ means all quantum corrections. For a relevant deformation, the quantum effect becomes subdominant in the UV region. In general, the $\b$-function is given by a function of the coupling constant relying on the RG scale. Then, the vev of the operator is derived from the variation of the generating functional with respect to the coupling constant, as shown in \eq{Relation:vevfromG}. For a marginal deformation with $\D=d$, the classical $\b$-function vanishes and the action still preserves the scale invariance. Therefore, the classical RG flow becomes trivial and the trace of the stress tensor vanishes at the classical level. However, this is not the case at the quantum level. The quantum effect allows a nonvanishing $\b$-function. If $\b_q = 0$, for example, the corresponding operator is called a truly marginal operator. In this case, the conformal symmetry of the UV theory preserves under the RG transformation even at the quantum level. Therefore, the deformed theory remains as the CFT in the entire RG scale. For $\b_q < 0$ (or $\b_q > 0$), we call such an operator a marginally relevant (or marginally irrelevant) operator. A marginally relevant (or irrelevant) operator behaves similar to a relevant (or irrelevant) operator except that a marginally irrelevant operator unlike an irrelevant operator still allows a UV fixed point. The marginally relevant or irrelevant deformation involves the nontrivial quantum correction and give rise to a nonvanishing trace of the stress tensor which we called the trace anomaly.

\section{Holographic dual of a marginal deformation}

According to the AdS/CFT correspondence, a $d$-dimensional QFT is dual to a classical $(d+1)$-dimensional gravity theory. Therefore, we may expect that a gravity theory realizes the RG flow of a  dual QFT. If possible, the AdS/CFT correspondence would be helpful to understand the nonperturbative RG flow because the AdS/CFT correspondence claims that the dual gravity maps to a QFT involving all quantum effects. Now, we investigate how we can describe the RG flow including quantum effects on the dual gravity side. In this work, we concentrate on a gravity theory which is dual to a UV CFT deformed by a classically marginal operator. Although the microscopic details of the deformed theory are obscure, studying the dual gravity theory can give us important information about the RG flow of the deformed QFT.

From now on, we focus on the case of $d=4$. According to the AdS/CFT correspondence, the mass of the bulk field is related to the conformal dimension of the dual operator 
\be
\D = 2 +\sqrt{4 + \fr{m^2}{R^2}} .
\ee
If we take into account a massless scalar field with $m=0$, the dual operator corresponds to a marginal operator with a conformal dimension $\D=4$. To consider a classically marginal deformation, we start with the following Euclidean Einstein-scalar gravity
\be			\la{act:EuclEins-scalar}
S =  - \fr{1}{2 \k^2} \int d^{5} X \sqrt{g} \ls {\cal R}  - 2 \L - \fr{1}{2} g^{MN}
\pa_M \ph \pa_N \ph  \rs + \fr{1}{\k^2} \int_{\pa {\cal M}} d^4 x \sqrt{\g} \ K ,
\ee
where $\L=-6/R^2$ is a cosmological constant with an AdS radius $R$. Here, $g_{MN}$ and $\g_{\m\n}$ indicate a bulk metric and an induced metric on the boundary respectively. Since the variation of the gravity action usually has a radial derivative of the metric at the boundary, the last term called the Gibbons-Hawking term is required to get rid of such a radial derivative term. Assuming that the scalar field depends only on the radial coordinate and that the boundary space is flat, the most general metric ansatz preserving the boundary's planar symmetry is expressed as
\be		\la{met:normalcoord}
ds^2 = e^{2 A(y)} \d_{\m\n} d x^{\m} dx^{\n}  + dy^2 .
\ee

The detail of this geometry is governed by $\ph(y)$ and $A(y)$ satisfying the following equations of motion
\be
0 &=& 24 \dot{A}^2 - \dot{\ph}^2 + 4 \L , 	\la{eq:consteq} \\
0 &=& 12 \ddot{A} + 24 \dot{A}^2 +   \dot{\ph}^2 + 4 \L  \la{eq:Adynamics} , \\
0 &=& \ddot{\ph} + 4 \dot{A} \dot{\ph}  ,  \la{eq:phdynamics}
\ee
where the dot means a derivative with respect to $y$. Here the first equation is a constraint, while the others describe dynamics of $A$ and $\ph$. Note that only two of the above three equations are independent. These equations allow the following analytic solution   \cite{Csaki:2006ji,Kim:2007qk}
\be		 
\ph &=&  \ph_0 + \et {\sqrt\fr{3}{2}} \log \ls \fr{4 \sqrt{6} - \ph_1 z^4/R^4 }{4 \sqrt{6} 
+  \ph_1 z^4/R^4 } \rs ,  \la{sol:ESeq} \\
e^{2 A(y)} &=& \fr{R^2}{z^2} \sqrt{1-  \fr {\et^2 \ph_1^2}{96} \fr{z^8}{R^8}} \la{Solution:gmetric} ,
\ee
with
\be			\la{met:bounarymetric}
z  = R e^{- y/R} ,
\ee
where $\ph_0$ and $\ph_1$ are two integral constants.  Assuming that two integral constants are positive, $\ph_0$ maps to a coupling constant of the dual UV CFT defined at the boundary ($y=\inf$). The invariance of the gravity action under the parity, $\ph \to - \ph$, allows two different profiles with $\et=\pm 1$. According to the AdS/CFT correspondence, a classical gravity theory is matched to a full quantum theory defined at the boundary.  Therefore, the classical geometric solution \eq{sol:ESeq} can give us information about the quantum effect of the deformation. Since the inner geometry of \eq{sol:ESeq} differs from the AdS space, the conformal symmetry is broken in the IR regime. This fact indicates that the deformation is not truly marginal. In other words, the deformation in \eq{sol:ESeq} is either marginally relevant or marginally irrelevant due to the quantum corrections. 

There exists another solution which yields a truly marginal deformation. When the scalar field has a constant profile which corresponds to the case with $\et=0$ in \eq{sol:ESeq}, its gravitational backreaction is absent. Thus, the AdS metric of the undeformed theory still becomes the solution of the deformed theory 
\be		 \la{Solution:trivial}
\ph &=&  \ph_0 ,  \\
A(y) &=& \fr{y}{R} .
\ee
Since the resulting geometry is still the AdS space, the deformed theory remains as a CFT, on the dual field theory side. In this case, the undeformed and deformed theories are different because their dual CFTs have different coupling constants. This is a typical feature of a truly marginal deformation which shifts the value of the coupling constant with a vanishing $\b$-function.

We have several important remarks. The classical conformal dimension of the dual operator can be easily read from the profile of the bulk scalar field. Near the boundary, for example,  a general massive bulk scalar field expands into
\be			\la{Solution:scalarfield}
\ph = c_0 z^{4-\D} + \ c_1  z^{\D}+ \cdots  .
\ee
Recalling that the radial coordinate maps to the energy scale of the dual field theory, the scaling behavior of the scalar field, without an appropriate renormalization procedure, allows us to identify $c_0$ and $c_1$ with a source (or coupling constant) and the vev of a dual operator, respectively. This identification is consistent with the classical scaling dimension of the dual QFT. If we further take into account quantum corrections through an appropriate renormalization, the classical scaling dimension modifies along the RG flow. In the above gravity action, the bulk scalar field is dimensionless because the gravitational constant has the dimension of $mass^{-3}$. Therefore, the bulk field is matched to a dimensionless coupling constant of the dual QFT. 

\subsection{Holographic description of the RG flow}

The RG equation is usually represented as first-order differential equations, while the equations of motion determining the dual geometry are governed by the second-order differential equations. To derive the RG flow from the bulk equations, we reformulate the bulk equations by applying the Hamilton-Jacobi formalism in which the bulk equations are rewritten as the first-order differential equations \cite{deBoer:1999tgo,Verlinde:1999xm,Skenderis:1999mm,Boer2001,Papadimitriou:2010as,Papadimitriou:2004ap}. To do so, we first notice that the metric solution in \eq{met:normalcoord} is a specific case of the following general metric
\be			\la{met:admdecomp}
ds^2  
&=& N^2 dy^2  + \g_{\m\n}(x,y)  dx^{\m}   dx^{\n}  ,
\ee
where $N$ is a lapse function and $\g_{\m\n} = e^{2 A(y)} \d_{\m\n}$. Note that the lapse function is non-dynamical and that varying the action with respect to the lapse function gives rise to a constraint equation. Therefore, we can set $N=1$, without loss of generality, after all calculation.

Regarding the radial coordinate $y$ as a Euclidean time, we can rewrite the previous Einstein-scalar gravity action as a functional form of the extrinsic curvature \cite{deBoer:1999tgo,Verlinde:1999xm,Boer2001}
\be
S = \int d^{4} x d y \sqrt{g} \ {\cal L}  ,
\ee
with
\be
{\cal L}  =  \fr{1}{2 \k^2} \lb N \ls - {\cal R}^{(4)} +K_{\m\n} K^{\m\n} - K^2 + 2 \L  \rs
+ \fr{1}{2N}    \dot{\ph}^2   \rb ,
\ee
where the extrinsic curvature tensor is defined as
\be
K_{\m\n} = \fr{1}{2 N} \fr{\pa \g_{\m\n}}{\pa y} .
\ee
Above $ {\cal R}^{(4)}$ denotes an intrinsic curvature of the boundary spacetime. Since the boundary is flat in our setup, ${\cal R}^{(4)}$ automatically vanishes. The canonical momenta of the boundary metric $\g_{\m\n}$ and scalar field $\ph$ are given by 
\be			\la{def:conjugatemomenta}
\pi_{\m\n} &\equiv& \fr{\pa S}{\pa \dot{\g}^{\m\n}}
= - \fr{1}{2 \k^2}  \ls K_{\m\n} - \g_{\m\n} K \rs , \nn
\pi_{\ph} &\equiv& \fr{\pa S}{\pa \dot{\ph}}
= \fr{1}{2  \k^2} \dot{\ph}   .
\ee
These canonical momenta enable us to reexpress the action as the first-order form
\be
S= \int d^{4} x d y \sqrt{g} \ \ls \pi_{\m\n} \dot{\g}^{\m\n} + \pi_{\ph} \dot{\ph}  - N {\cal H}
\rs ,
\ee
where the Hamiltonian density is given by
\be
{\cal H} = 2 \k^2 \ls \g^{\m\r} \g^{\n\s} \pi_{\m\n} \pi_{\r\s} - \fr{1}{3} \pi^2 + \fr{1}{2} \pi_{\ph}^2 \rs  - \fr{\L}{\k^2}    .
\ee
with $\pi = \g^{\m\n} \pi_{\m\n}$. The variation of this action with respect to the lapse function leads to the Hamiltonian constraint, ${\cal H}=0$. This Hamiltonian is a generator of the translation in the $y$-direction. All solutions connected by this transformation are gauge equivalent.

After imposing the Hamiltonian constraint, the variation of the action finally results in the variation of the boundary action
\be
\d S_B =  \int_{\pa {\cal M}} d^4 x \sqrt{\g} \ls \pi_{\m\n}  \d {\g}^{\m\n} + \pi_{\ph} \d {\ph} \rs ,
\ee
where all variables are defined at the boundary. In the Hamilton-Jacobi formalism, the canonical momenta are defined as
\be			\la{def2:conmon}
\pi_{\m\n} =   \fr{1}{\sqrt{\g}} \fr{\d S_B}{\d \g^{\m\n}} \quad {\rm and} \quad
\pi_{\ph} =  \fr{1}{\sqrt{\g}} \fr{\d S_B}{\d \ph} .
\ee
According to the AdS/CFT correspondence, we identify the above boundary action with the generating functional of the dual QFT. We also identify the radial coordinate of the bulk geometry with the energy scale of the dual QFT. Then, we can investigate the RG flow of the dual QFT by changing the boundary position on the gravity side. To map the above equations to the RG equations correctly, we need to resolve two issues. First, we have to clarify how the radial position of the boundary is related to the RG scale \cite{Kim:2016hig,Kim:2016ayz,Kim:2017lyx}. Second, the above boundary action is an unrenormalized generating functional because we did not get rid of UV divergences yet. Therefore, we have to renormalize the above boundary action by adding appropriate counterterms.

Although the RG flow can vary the value of coupling constants including the metric, it does not change the background geometry where the QFT is defined. On the dual gravity side, this implies that the boundary spacetime described by $ds^2 = \g_{\m\n} dx^\m dx^\n$ must be invariant under the scale transformation. Since the metric component in \eq{met:admdecomp} scales by $e^{A(\bar{y})} \to  e^\s  e^{A(\bar{y} )}$ under $x \to e^{-\s} x$ or $\m \to e^{\s} \m$, we can associate the RG scale of the dual QFT with the metric component at the boundary
\be         \la{Result:RGscale}
\m=\fr{e^{A (\bar{y})}}{R} ,
\ee
where $A (\bar{y})$ indicates the value of $A$ at the boundary. Above, we introduced the AdS radius $R$ for the correct dimension count. This relation represents how the RG scale changes when the boundary moves in the radial direction.  From now on, we concentrate on the dual QFT defined at the boundary, so we drop out the bar symbol. For $\et=\pm 1$, the RG scale of the dual QFT is related to the radial position of the boundary by    
\be			\la{Result:RGscaley}
\m 
= \fr{e^{y/R} }{R} \ls 1- \fr{\ph_1^2 }{96} e^{- 8y/R}\rs^{1/4} .
\ee

Even though we know how to relate the radial coordinate to the RG scale, we need to add  appropriate counterterms to remove the UV divergences appearing in the boundary action. In this renormalization procedure, there can exist several different counterterms which have the same UV divergent terms \cite{Park:2013dqa}. Although the regular terms do not give any effect on the renormalization, they can provide a nontrivial effect on the RG flow in the intermediate energy scale. The regular part of the counterterms is associated with the quantum corrections and seriously modify the IR physics. Therefore, it is an important to fix the counterterms correctly to understand the RG flow and its IR physics. The UV divergence of the above boundary action appears at a UV cutoff $\bar{y} \to \inf$ due to the invariant integral measure
\be
\int d^4 x \sqrt{\g} \sim \int d^4 x \, e^{4 \bar{y}/R} .
\ee
On the other hand, the integrands of the boundary action in the aymptotic region behave as
\be
\pi_{\m\n} \g^{\m\n} &\sim& \dot{A} \sim \fr{1}{R} , \nn
\pi_{\ph} \ph &\sim&  e^{-4 \bar{y}/R} .
\ee
As shown in these asymptotic behaviors, the gravity part, $\int \sqrt{\g} \, \pi_{\m\n} \g^{\m\n}$, gives rise to a UV divergence proportional to $e^{4 y/R}$, whereas the scalar field part, $ \int \sqrt{\g} \, \pi_{\ph} \ph$, does not make any additional UV divergence. Therefore, we need the counterterm which cancels only the divergence of the gravity part. In other words, since the marginal deformation does not any additional UV divergence, we exploit the same counterterm used in the undeformed CFT \cite{Balasubramanian:1999re}
\be
S_{ct} = - \fr{1}{2 \k^2} \int d^4 x  \sqrt{\g} \ {\cal L}_{ct}  .
\ee
with
\be             \la{Result:ctformarginal}
{\cal L}_{ct} = \fr{6 }{ R}  .
\ee
Then, the renormalized action is given by
\be
\G [\g_{\m\n},\ph; \bar{y}]= S_B - S_{ct} .
\ee

The resulting renormalized action is finite and corresponds to the renormalized generating functional of the dual QFT. The UV cutoff we introduced is artificial, so the the physical renormalized action must be independent of this artificial UV cutoff. The scale independence of the renormalized action leads to the RG equation 
\be
0 
= \m \fr{\pa \G}{\pa \m}  + \fr{\pa \g^{\m\n}}{\pa \log \m}  \fr{\pa \G}{\pa \g^{\m\n}} +  \fr{\pa \ph}{\pa \log \m}  \fr{\pa \G}{\pa \ph}  ,
\ee
where $\m$ is the RG scale satisfying \eq{Result:RGscale}. Here $\ph$ indicates the boundary value of the bulk scalar field. Identifying $\ph$ with the dimensionless coupling  of the dual QFT, the vev of a scalar operator is derived from the generating functional, as it should do. Due to the following relation
\be
 \fr{d \g^{\m\n}}{d \log \m}  = - 2 \g^{\m\n} .
\ee
we further rewrite the above RG equation as the usual form
\be		\la{res:hRGeq}
0 = \fr{\m}{\sqrt{\g}} \fr{\pa \G}{\pa \m}  + \g^{\m\n} \bra T_{\m\n} \ket + \b_\ph  \bra O \ket  ,
\ee
with the following definitions
\be
\b_\ph  &\equiv&  \fr{\pa \ph}{\pa \log \m} ,\\
\bra T_{\m\n} \ket &\equiv& - \fr{2}{\sqrt{\g}} \fr{\pa \G}{\pa \g^{\m\n} }  =  -  \ls 2 \pi_{\m\n} - \fr{1}{2 \k^2} \g_{\m\n} {\cal L}_{ct} \rs , \\
\bra  O \ket &\equiv&    \fr{1}{\sqrt{\g}}  \fr{ \d \G}{\d \ph}  = \pi_\ph  + \fr{1}{2 \k^2} \fr{\pa {\cal L}_{ct} }{\pa \ph}  .  \la{Result:vevofoperator}
\ee
This is the same as the generalized RG equation \eq{Result:gRGeq} of a QFT, where we take into account the metric as an additional coupling constant.

\subsection{RG flow triggered by a marginal operator}

When we describe the holographic RG flow, it is more convenient to introduce a superpotential. To do so, we return to the bulk equations of motion. Since the bulk equations of motion,\eq{eq:consteq}, \eq{eq:Adynamics}, and \eq{eq:phdynamics}, are depending only on $\dot{A}$ and  $\ddot{A}$, we can introduce a superpotential which satisfies \cite{Freedman:1999gp,Skenderis:1999mm,DeWolfe:1999cp,Csaki:2000wz, Gubser:1999pk, Kehagias:1999tr,Csaki:2006ji}
\be		\la{ans:superpotential}
W(\ph) = 6  \dot{A}  .
\ee
Then, the bulk equations reduce to two first-order differential equations
\be
2 \L   &=& \fr{1}{2} \ls \fr{\pa W  }{\pa \ph} \rs^2 - \fr{1}{3} W^2  , \la{res:Esequaiton1}  \\
\dot{\ph} &=& - \fr{\pa W  }{\pa \ph}   . \la{res:Esequaiton2} 
\ee
Here the first equation is just the Hamiltonian constraint which determines the superpotential as a function of $\ph$. Then, we rewrite the RG equation in terms of the superpotential
\be		
0 = \fr{\m}{\sqrt{\g}} \fr{\pa \G}{\pa \m}  + \g^{\m\n} \bra T_{\m\n} \ket + \b_\ph  \bra O \ket  ,
\ee
with 
\be
\b_\ph  &=& - \fr{6 }{W} \fr{\pa W  }{\pa \ph} , \\
\bra T_{\m\n} \ket &=&   \fr{1}{\k^2}  \ls K_{\m\n}  - \g_{\m\n} K \rs-  \fr{3}{\k^2 R}  \g_{\m\n}  , \\
\bra  O \ket &=& \fr{1}{2  \k^2}  \fr{\pa W  }{\pa \ph} .   
\ee
For a marginal deformation, $\pa {\cal L}_{ct}/\pa \ph$ automatically vanishes because $ {\cal L}_{ct}$ is given by a constant, as shown in \eq{Result:ctformarginal}.

\subsubsection{Marginally relevant or irrelevant deformation}

The Hamiltonian constraint \eq{res:Esequaiton1} allows two different types of solutions. 
The first one is given by \cite{Papadimitriou:2010as,Papadimitriou:2004ap,Csaki:2006ji,Kim:2007qk}
\be		\la{res:1counterterm}
W =   \fr{6 }{R}  \cosh \ls \sqrt{\fr{2}{3}} \ls \ph - \ph_0 \rs \rs .
\ee
After substituting the superpotential into \eq{res:Esequaiton2} and solving \eq{res:Esequaiton2}, this superpotential reproduces
the solution in \eq{sol:ESeq} and \eq{Solution:gmetric} with $\et = \pm1$. From the profile of the bulk scalar field \eq{sol:ESeq}, we can easily read off the corresponding $\b$-function in the UV region 
\be         \la{Result:betafofphi}
\b_\ph  
= \fr{\et \ph_1}{\m^4} + {\cal O} \ls \fr{1}{\m^{12}}\rs .
\ee
Recalling that the $\b$-function of the marginal operator vanishes at the classical level, this nonvanishing $\b$-function comes thoroughly from the quantum effect. If we identify the bulk scalar field with the coupling constant of the dual QFT, the nonvanishing $\b$-function implies that the deformation operator is marginally relevant ($\b_\ph  < 0$) for $\et =  -1$ or marginally irrelevant ($\b_\ph  > 0$) for $\et = 1$, see Fig. 1. On the other hand, if we take the bulk scalar as the inverse of the coupling constant as will be seen in the next section, the deformation becomes marginally relevant for $\et=1$ and margiallry irrelevant for $\et=-1$. We see from \eq{Result:vevofoperator} that the vev of the deformation operator reduces to
\be		\la{eq:constraintcounter}
\bra O \ket  =  \frac{\sqrt{6} }{\kappa ^2 R}   \sinh \left(\sqrt{\frac{2}{3}} \left(\phi -\phi _0\right)\right) .
\ee
These nonvanishing $\b$-function and vev of the operator give rise to a nontrivial trace anomaly along the RG flow
\be		
\bra {T^\m}_{\m} \ket  = - \fr{\m}{\sqrt{\g}} \fr{\pa \G}{\pa \m}  - \b_\ph  \bra O \ket  .
\ee
This is the expected RG flow when the dual QFT deforms by a marginally relevant or irrelevant operator.

\subsubsection{Truly marginal deformation}

There exists another solution satisfying the Hamiltonian constraint. The second superpotential is given by
\be
W = \fr{6}{R} .
\ee
This reproduces the trivial solution, \eq{sol:ESeq} with $\et=0$. Plugging this superpotential into \eq{res:Esequaiton2} and \eq{ans:superpotential}, we finally reobtain the trivial solution in \eq{Solution:trivial}. In this case, the coupling constant $\ph=0$ of the undeformed CFT changes into $\ph=\ph_0$ for the deformed CFT. This is a typical feature of the truly marginal deformation satisfying 
\be
\b_\ph =\bra {T^\m}_\m \ket=0.
\ee 
Therefore, this truly marginal operator changes a CFT into another CFT with shifting the value of the coupling constant and without generating a nontrivial RG flow. 

\begin{figure}
\begin{center}
\vspace{-0cm}
\hspace{-0cm}
\includegraphics[angle=0,width=0.8\textwidth]{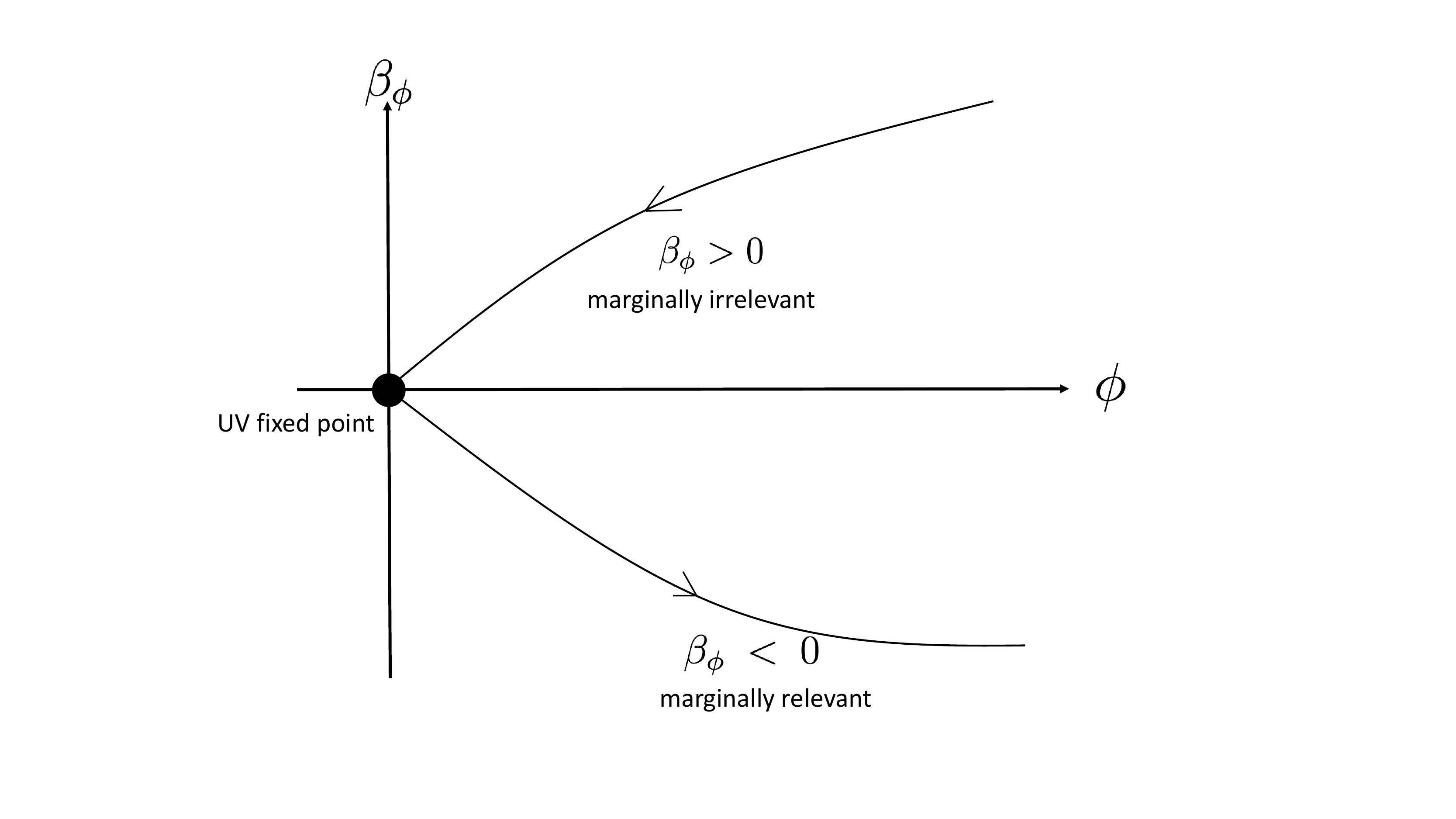}
\vspace{-0.5cm}
\caption{\small The RG flows caused by  marginally relevant and irrelevant operators.}
\label{number}
\end{center}
\end{figure}

\section{Holographic dual of the gluon condensation}

Several distinct condensations appear in QCD. They are usually associated with a certain spontaneous symmetry breaking and responsible for the mass of hadrons. Due to the Lorentz invariance, the condensation should be a Lorentz scalar and have vanishing charges under various global and local symmetries. The well-known condensations in QCD are the chiral and gluon condensations. The chiral condensation is the condensate of two fermions with breaking the chiral symmetry spontaneously and yields a large effective mass to quarks and most hadrons. The other example is the gluon condensation. For a four-dimensional Yang-Mills theory, the kinetic term of the gauge field is given by
\be
S_{YM} = - \fr{1}{4  g_{YM}^2}  \int d^4 x \sqrt{\g}   \  \  \Tr  F^2  .
\ee
In this case, since the Yang-Mills coupling constant is classically marginal, the classical $\b$-function automatically vanishes. However, the one-loop quantum correction generates a nonvanishing $\b$-function  \cite{Wilson:1973jj,Gross:1973ju,tHooft:1998qmr}.

When QCD deforms by the gluon condensation, we are able to regard $\ph=1/(4  g_{YM}^2)$ as a coupling constant and $G = -  \Tr  F^2$ as a deformation operator called the gluon condensation, respectively. Although it is not clear whether the gluon condensation is associated with the known phase change, there are many indications for the nonvanishing gluon condensation in lattice QCD simulations. The gluon condensation may be partly responsible for masses of hadrons and  leads to a nontrivial RG flow closely related to the trace anomaly. The lattice QCD simulations expects that the one-loop quantum correction in the UV region leads to the following trace anomaly \cite{DiGiacomo:1981lcx,Trinchero:2013joa}
\be
\bra {T^{\m}}_{\m} \ket = - \fr{N_c}{8 \pi } \fr{\b_\l}{\l^2} \bra G \ket ,   \la{Result:GRGfloweqQCD}
\ee
where $N_c$ indicates the rank of the gauge group and $\b_\l$ means a $\b$-function of the 't Hooft coupling constant $\l$. In general, the condensation plays a crucial role in nonperturbative phenomena. To understand such nonperturbative features,  the AdS/CFT correspondence may be helpful. The effect of the gluon condensation on the trace anomaly can be understood by the holographic RG flow discussed before. Although the gravity model we considered is too simple to study IR physics, the present model is still useful to account for the UV physics because it gives rise to the leading behavior of the marginal deformation at least in the UV region.

Now, we specify the parameters of the gravity theory in terms of those of the dual QFT.  For applying the AdS/CFT correspondence, we define the `t Hooft coupling constant as $\l = N_c \, g_{YM}^2$ and take the double scaling limit where $N_c \to \infty$ and $g_{YM}^2 \to 0$ with a fixed $\l$. Despite an infinite rank of the gauge group, the AdS/CFT correspondence still has an advantage in catching the important feature of the nonperturbative RG flow. If we identify the vev of the dual operator with the gluon condensation, $\bra  G \ket$, the bulk scalar field is associated with the 't Hooft coupling constant 
\be
\ph = \fr{N_c}{ 4 \l}  .			\la{Relation:philambda}
\ee
In the large 't Hooft coupling constant limit ($\l \gg 1$), the leading behavior of $\ph$ is proportional to $1/\m^4$ in \eq{sol:ESeq}. This implies that the 't Hooft coupling constant is related to the RG scale by $\l \sim \m^4$ in the UV region ($\m \gg 1$). The $\b$-function \eq{Result:betafofphi} derived from the bulk scalar field can be reexpressed as the $\b$-function of the 't Hooft coupling constant 
\be			\la{Result:betabulk}
\b_\ph   
= -  \fr{N_c}{4  } \fr{\b_\l}{\l^2}   .
\ee
When we identify the bulk scalar field with the inverse of the 't Hooft coupling constant in \eq{Relation:philambda}, we have to take $\et=1$ rather than $\et=-1$ for describing a marginally relevant deformation. For example, if the rank of the gauge group $N_c$ is sufficiently larger than the number of other matter fields, $\b_\l $ has a negative value representing a marginally relevant deformation. In \eq{Result:betabulk}, the negative $\b_\l $ enforces a positive $\b_\ph$ which is the same as choosing $\et=+1$ in \eq{Result:betafofphi}. This is also consistent with our previous prescription that, when the bulk field is identified with the inverse of the coupling constant, $\et=+1$ describes a marginally relevant deformation.

From the holographic RG flow description studied before, the $\b$-function and the gluon condensation in the UV region are given by functions of the RG scale 
\be
\b_{\ph} &=&  \fr{\ph_1}{R^4} \fr{1}{\m^4} - \fr{\ph_1^3}{48 R^{12}} \fr{1}{\m^{12}}  + {\cal O} \ls \m^{-20} \rs         ,   \nn
\bra G \ket  &=& - \fr{\ph_1}{2 \k^2 R^5} \fr{1}{\m^4}  + {\cal O} \ls \m^{-28} \rs     .
\ee
Here  the $\b_\l$ and $\bra G \ket$ are also represented as functions of the 't Hooft coupling constant, for example, $\b_\l \sim -  \l$ and $\bra G \ket \sim - 1/\l$ in the UV region. The leading behavior of the gluon condensation, $\bra G \ket \sim  \m^{-4}$, is the form expected by the classical dimension counting in \eq{Result:scbehavior}. This result shows that the gluon condensation rapidly suppresses as the 't Hooft coupling constant increases. From \eq{Result:vevofoperator}, moreover, we show that the trace anomaly has the following RG scale dependence 
\be
\bra {T^{\m}}_{\m}\ket = - \fr{\ph_1^2}{4 \k^2 R^9} \fr{1}{\m^8} + \frac{\phi _1^4}{384 \kappa ^2   R^{17}}  \fr{1}{\m^{16}}+ {\cal O} \ls \m^{-24} \rs    .
\ee
Comparing the results of the holographic RG flow, we find that the gluon condensation satisfies the following relation
\be			\la{Result:GRGfloweq}
\bra {T^{\m}}_{\m} \ket   =  - \fr{N_c }{8} \fr{\b_\l}{\l^2} \bra G \ket + {\cal O} \ls \l^{-4} \rs.
\ee
This is the form of the trace anomaly expected in the lattice QCD \cite{DiGiacomo:1981lcx,Trinchero:2013joa}. There are remarkable points in this holographic RG flow. Intriguingly, the holographic RG flow shows that the trace anomaly \eq{Result:GRGfloweqQCD} expected from the lattice QCD is valid up to the $\l^{-4}$ order. The holographic description of the gluon condensation leads to the expected behavior of the RG flow. The gluon condensation and the trace anomaly, as expected, rapidly suppress in the UV region, so that the conformal symmetry is restored at the UV fixed point.


\section{Discussion}

In this work, we studied the holographic RG flow of a CFT deformed by a marginal operator. We discussed how we can understand the RG flow of a boundary QFT in terms of the Hamilton-Jacobi formulation on the dual gravity side. At the classical level, a marginal operator does not change the CFT. The quantum effect, however, can lead to the nontrivial modification of the CFT along the RG flow. Using the holographic description, we studied how the quantum effect of the marginal deformation modifies the $\b$-function and the vev of the operator. Furthermore, we compared this result with the result of the gluon condensation known in QCD.

There were several distinct prescriptions to realize the nonperturbative RG flow of the QFT on the dual gravity side. In the present work, we exploited the Hamilton-Jacobi formalism which allows us to rewrite the gravity equations as the first-order differential equations. The RG equations are generally given by the first-order differential equations, so that the Hamilton-Jacobi formulation is useful to understand the RG flow of the dual QFT. When we applied the Hamilton-Jacobi formulation, it suffers from the UV divergences similar to those appearing in the QFT renormalization. We discussed the counterterms, which get rid of the UV divergences of the holographic renormalization, and then find the finite boundary action, which is identified with the generating functional of the dual QFT.

After the holographic renormalization, we explicitly showed that the quantum correction modifies a classically marginal operator into one of the truly marginal, marginally relevant and marginally irrelevant operators at the quantum level. More precisely, the quantum correction gives rise to a nonvanishing $\b$-function for marginally relevant and irrelevant deformations. If we focus on the marginally relevant deformation, the undeformed theory at the UV fixed point becomes unstable under this marginally relevant deformation. Thus, the UV CFT flows to a new IR theory. In this case, other quantities like the stress tensor and the vev of the operator also vary along the RG flow. We explicitly calculated the RG scale dependence of these quantities near the UV fixed point. We also showed that the holographic RG flow reproduces the known trace anomaly of the gluon condensation in the lattice QCD \cite{DiGiacomo:1981lcx,Trinchero:2013joa}. Intriguingly, the holographic RG flow indicates that this trace anomaly is valid only up to $\l^{-4}$ order in the UV region.

In the present work, the Einstein-Scalar gravity we considered has no well defined IR geometry because of the existence of a singularity at $z^4 = 4 \sqrt{6}/\ph_1$. This means that the dual QFT of the present model is IR incomplete, so that the end of the RG flow is not manifest. This means that the present model is not valid in the IR region. Nevertheless, the results we obtained are still valid in explaining the quantum correction in the UV region because it gives rise to the leading contribution. To study the IR physics further, we have to consider a more general gravity theory which allows a well-defined IR fixed point. For example, we can take into account a scalar field potential with higher-order terms  
\be
V(\ph) = \sum_{n \ge 3} a_n \, \ph^n .
\ee
For a relevant deformation, this potential crucially modifies IR physics and can allow an IR fixed point. Despite this fact, its contribution in the UV region is subdominant. Due to this reason, in the present work we just focus on the UV behavior of the RG flow.  Nevertheless, it is still important to take into account higher-order terms to understand the IR phenomena. We hope to report more results on this issue in future works. \\

\vspace{0.5cm}

{\bf Acknowledgement}

This work was supported by the National Research Foundation of Korea(NRF) grant funded by the Korea government(MSIT) (No. NRF-2019R1A2C1006639).

\vspace{0.5cm}


\bibliographystyle{apsrev4-1}
\bibliography{References}

\end{document}